\documentclass[preprint,aps]{revtex4-1}
\usepackage{graphicx}
\usepackage{amsmath}
\usepackage{subfigure}

\newcommand{\be}{\begin{equation}}
\newcommand{\ee}{\end{equation}}
\newcommand{\ben}{\begin{eqnarray}}
\newcommand{\een}{\end{eqnarray}}

\newcommand{\nd}{\noindent}

\begin{document}

\title{Quantum correlations in the thermodynamic limit: the XY-model }
\author{J. Batle$^{1}$,  A. Plastino$^{2}$,  A.R. Plastino$^{3,\,4}$, M. Casas$^{1}$}
\affiliation{ $^1$Departament de F\'{\i}sica and IFISC, Universitat de
les Illes Balears,
 07122 Palma de Mallorca, Spain  \\\\
$^2$IFLP-CCT-CONICET, National University La Plata,
  C.C. 727, 1900 La Plata, Argentina  \\\\ $^3$CREG-UNLP-CONICET,
 National University la Plata, C.C. 727, 1900 La Plata, Argentina
\\\\ $^4$Instituto Carlos I de F\'{\i}sica Te\'{o}rica
 y Computacional, Universidad de Granada, Granada, Spain}
\date{\today}

\begin{abstract}

We investigate thermal properties of quantum correlations in
the thermodynamic limit with reference to the XY-model

\vskip 2cm

\end{abstract}


\maketitle

\section{Introduction}

\nd
 Since the formalization by Werner \cite{W89} of the modern concept
of quantum entanglement it has become  clear that there exist
entangled states that comply with all Bell inequalities (BI). This
entails that non-locality, associated to BI-violation, constitutes
a  non-classicality manifestation exhibited only by just a subset
of the full set of states endowed with quantum correlations. Later
work by Zurek and Ollivier \cite{olli} established that not even
entanglement captures all aspects of quantum correlations. These
authors introduced an information-theoretical measure, quantum
discord, that corresponds to a new facet of the ``quantumness"
that arises even for non-entangled states. Indeed, it turned out
that the vast majority of quantum states exhibit a finite amount
of quantum discord.

\nd The tripod non-locality-entanglement-quantum discord is of
obvious interest and possesses  technological implications. The
crucial role played by quantum entanglement in quantum information
technologies is well known \cite{Nielsen}. In some cases, however,
entangled states are useful to solve a problem if and only if they
violate a Bell inequality \cite{comcomplex}. Moreover, there are
important instances of non-classical information tasks that are
based directly upon non-locality, with no explicit reference to
the quantum mechanical formalism or to the associated concept of
entanglement \cite{device}. Last, but certainly not least, recent
research indicates that quantum discord is also a valuable
resource for the implementation of non-classical information
processing protocols \cite{geom,ferraro,dattaprl,luo,datta}. On
the light of these developments, it becomes imperative to conduct
a systematic exploration of the connections between the tripod
members.

\nd We investigate here the relation between quantum discord and
entanglement and in an infinite system, namely, the XY model in
the thermodynamic limit \cite{LSM}. This model, like the
celebrated Ising and Heisenberg models, is one of the paradigmatic
 systems in statistical mechanics. The Hamiltonian of the
anisotropic one-dimensional spin-$\frac{1}{2}$ XY model in a
transverse magnetic field $h$ ($N$ particles) reads

\begin{equation} \label{equacioXY}
H=\sum_{j=1}^N [ (1+\gamma)S_x^{j}S_x^{j+1} + (1-\gamma)S_y^{j}S_y^{j+1} ]-h\sum_{j=1}^N S_z^{j},
\end{equation}
\noindent where $\sigma^j_u=2S_u^{j}$ ($u=x,y,z$) are the Pauli
spin-$\frac{1}{2}$ operators on site $j$, $\gamma \in [0,1]$ and
$\sigma^{j+N}_u=\sigma^j_u$. The  model (\ref{equacioXY}) for
$N=\infty$ is completely solved by applying a Jordan-Wigner
transformation \cite{LSM,Barouch}, which maps the Pauli (spin 1/2)
algebra into canonical (spinless) fermions. The system (except for
the isotropic case $\gamma=0$) undergoes a
paramagnetic-to-ferromagnetic quantum phase transition (QPT)
\cite{qpt,Sachdev} driven by the parameter $h$ at $h_c=1$ and
$T=0$. It is well known that near factorization a characteristic
length scale naturally emerges in the system, which is
specifically related with the entanglement properties and diverges
at the critical point of the fully isotropic model \cite{baroni}.

\nd It is thus our intention in this communication to  study the
interplay of entanglement and quantum discord for the XY model in
the thermodynamic limit. To such an effect we will consider in
Section II the correlations existing between a pair of qubits
located at two given sites. For comparison purposes we shall also
discuss in Section III the correlations between pairs of qubits in
the finite Heisemberg model. Some conclusions are drawn in Section
IV.

\subsection{Quantum discord}

\nd Quantum discord \cite{geom,olli} constitutes a quantitative
measure of the ``non-classicality" of bipartite correlations as
given by the discrepancy between the quantum counterparts of two
classically equivalent expressions for the mutual information.
More precisely, quantum discord is defined as the difference
between two ways of expressing (quantum mechanically) such an
important entropic quantifier. Let $\rho$ represent a state of a
bipartite quantum system consisting of two subsystems $A$ and $B$.
If $S(\rho)$ stands for the von Neumann entropy of matrix $\rho$
and
 $\rho_A$ amd $\rho_B$ are the reduced (``marginal") density matrices
describing the two subsystems, the quantum mutual information
(QMI) $M_q$ reads \cite{olli} \be \label{uno} M_q(\rho)= S(\rho_A)
+  S(\rho_B) - S(\rho). \ee This quantity is to be compared to
another quantity ${\tilde M}_q(\rho)$, expressed using conditional
entropies, that classically coincides with the mutual information.
To define ${\tilde M}_q(\rho)$ we need first to consider the
notion of conditional entropy. If a complete projective
measurement $\Pi_j^B$ is performed on B and (i) $p_i$ stands for
$Tr_{AB}\,\Pi_i^B\,\rho$ and (ii) $\rho_{A|\vert \Pi_i^B}$ for
$[\Pi_i^B\,\rho\,\Pi_i^B/p_i]$, then the conditional entropy
becomes

\be  \label{unobis}  S(A \vert\,\{ \Pi_j^B \}) =\sum_i\,p_i\,
S(\rho_{A|\vert \Pi_i^B}), \ee and ${\tilde M}_q(\rho)$ adopts the
appearance

\be \label{dos} {\tilde M}_q(\rho)_{\{ \Pi_j^B \}} = S(\rho_A)-
S(A \vert\,\{ \Pi_j^B \}). \ee Now, if we minimize over all
possible $ \Pi_j^B$ the difference $M_q(\rho)-{\tilde
M}_q(\rho)_{\{ \Pi_j^B \}}$ we obtain the quantum discord
$\Delta$, that quantifies non-classical correlations in a quantum
system, {\it  including those not captured by entanglement}. One
notes then that only states with zero $\Delta$ may exhibit
strictly classical correlations.
Among many valuable discord-related works we just mention two at this point that are intimately related to the present one, e.g., those of  Zambrini et al. \cite{zambrini} and
Batle et al. \cite{ours}.

\section{Two qubits in the infinite $XY$ model}

\nd The general two-site density matrix is expressed as

\begin{equation} \label{rho2}
\rho_{ij}^{(R)} = \frac{1}{4} \,
\Bigg[ I + \sum_{u,v} T_{uv}^{(R)}
\sigma^i_u \otimes \sigma^j_v  \Bigg].
\end{equation}
\noindent $R=j-i$  is the distance between spins, $\{u,v\}$ denote
any index of $\{\sigma_0,\sigma_x,\sigma_y,\sigma_z\}$, and
$T_{uv}^{(R)} \equiv \langle \sigma^i_u \otimes \sigma^j_v
\rangle$. Due to symmetry considerations, only
$\{T_{xx}^{(R)},T_{yy}^{(R)},T_{zz}^{(R)},T_{xy}^{(R)}\}$ do not
vanish. Barouch {\it et al} \cite{Barouch} have provided exact
expressions for two-point quantum correlations, together with details of
the dynamics associated with an external field $h(t)$. 
 For the purposes of this paper, we will
consider only systems which at time $t = 0$ are in
thermal equilibrium at temperature $T$. We have then the canonical ensemble expression $\rho(t=0)= \exp{[-\beta\,H]}$, where $\beta = 1/kT$ and $k$ is the Boltzmann constant.
     Following
\cite{Barouch}, one obtains
$T_{xx}^{(1)}=G_{-1},T_{yy}^{(1)}=G_{1},T_{zz}^{(1)}=G_0^2-G_1G_{-1}-S_1S_{-1}$,
and $T_{xy}^{(1)}=S_1$, where

\begin{eqnarray}
G_R&=& \frac{\gamma}{\pi} \int_0^{\pi} d\phi \sin (R\phi)
\frac{\tanh \big[ \frac{1}{2}\beta \Lambda(h_0)
\big]}{\Lambda(h_0)\Lambda^2(h_f)}\cr &&\times [ \gamma^2\sin
^2\phi + (h_0-\cos \phi)(h_f - \cos \phi) \cr
&&-(h_0-h_f)(h_f-\cos \phi) \cos (2\Lambda(h_f)t) ] \cr
&&-\frac{1}{\pi} \int_0^{\pi} d\phi \cos (R\phi) \frac{\tanh \big[
\frac{1}{2}\beta \Lambda(h_0)
\big]}{\Lambda(h_0)\Lambda^2(h_f)}\cr &&\times \big[ \{
\gamma^2\sin ^2\phi + (h_0-\cos \phi)(h_f - \cos \phi)\}(\cos \phi
- \cr && h_f)-(h_0-h_f)\gamma^2\sin ^2\phi \cos (2\Lambda(h_f)t)
],\\ \label{G} && \cr S_R&=&\frac{\gamma
(h_0-h_f)}{\pi}\int_0^{\pi} d\phi\sin (R\phi) \sin \phi \frac{\sin
[ 2\Lambda(h_f)t ]}{\Lambda(h_0)\Lambda(h_f)}, \label{GGG}
\end{eqnarray}
\noindent with $\Lambda(h)=[\gamma^2\sin^2\phi + (h-\cos
\phi)^2]^{1/2}$. $G_R$ is the two-point correlator appearing in
the pertinent Wick-calculations and $M_z=\frac{1}{2}G_0$. 
The two-spin correlation functions are given by \cite{Barouch}

\begin{align}
  \langle \sigma_x^i\sigma_x^{i+R}\rangle &=
  \begin{vmatrix} G_{-1} & G_{-2} & \cdots & G_{-R} & \\ G_0 & G_{-1}
    & \cdots & G_{-R+1} & \\ \vdots & \vdots & \ddots & \vdots & \\
    G_{R-2}& G_{R-3}& \cdots & G_{-1} &
  \end{vmatrix}, \label{eq:corr1} \\ \langle
  \sigma_y^i\sigma_y^{i+R}\rangle &= \begin{vmatrix} G_{1} & G_{0} &
    \cdots & G_{-R+2} & \\ G_2 & G_{1} & \cdots & G_{-R+3} & \\ \vdots
    & \vdots & \ddots & \vdots & \\ G_{R}& G_{R-1}& \cdots & G_{1} &
  \end{vmatrix},
  \label{eq:corr2} \\ \langle \sigma_z^i\sigma_z^{i+R}\rangle &=
  4\langle\sigma_z\rangle^2 - G_RG_{-R} \label{eq:corr3},
\end{align}

\noindent where $R=j-i$  (distance between spins). In the case where more than 
two particles are considered, the previous correlators no longer possess their previous Toeplitz matrix structure 
\cite{Batle2010}.


 It will prove convenient to cast the two qubit states
(\ref{rho2}) in two forms. States $\rho_{ij}^{(R)}$ are written in
the computational basis $\{
|00\rangle,|01\rangle,|10\rangle,|11\rangle \}$ as

\begin{equation} \label{rhocompu}
 \frac{1}{4} \left( \begin{array}{cccc}
1+4M_z+T_{zz} & 0 & 0 & T_{xx}-T_{yy}-i 2T_{xy} \\
0 & 1-T_{zz} & T_{xx}+T_{yy} & 0 \\
0 & T_{xx}+T_{yy} & 1-T_{zz} & 0 \\
T_{xx}-T_{yy}+i 2T_{xy} & 0 & 0 & 1-4M_z+T_{zz} \end{array} \right).
\end{equation}
\noindent These very states $\rho_{ij}^{(R)}$ acquire instead the
following form in the Bell basis $\{
|\Phi^{+}\rangle,|\Phi^{-}\rangle,|\Psi^{+}\rangle,|\Psi^{-}\rangle
\}$

\begin{equation} \label{rhoBell}
 \left( \begin{array}{cccc}
\rho_{11} & i\rho^{I}_{12} & i\rho^{I}_{13} & \rho^{R}_{14}\\
-i\rho^{I}_{12} & \rho_{22} & \rho^{R}_{23} & i\rho^{I}_{24}\\
-i\rho^{I}_{13} & \rho^{R}_{23} & \rho_{33} & i\rho^{I}_{34}\\
\rho^{R}_{14} & -i\rho^{I}_{24} & -i\rho^{I}_{34} & \rho_{44} \end{array} \right).
\end{equation}

\noindent This special form is such that one can use it to analytically compute the maximal violation of a Bell inequality,  a measure
for nonlocality \cite{Batle2010}.
 In turn states $\rho_{ij}^{(R)}$ in (\ref{rhocompu}) are of such
 special aspect that the quantum discord Qd turns out to be to be analytically given 
 (see Ref. \cite{rossig}). Nevertheless, the concomitant
Qd can be easily obtained, in different fashion, as follows. The most general
parameterization of the local measurement that can be implemented
on one qubit (let us call it B) is of the form $\{
\Pi_B^{0^{\prime}}=I_A \otimes |0^{\prime}\rangle \langle
0^{\prime}|, \Pi_B^{1^{\prime}}=I_A \otimes |1^{\prime}\rangle
\langle 1^{\prime}|\}$. More specifically we have

\begin{eqnarray} \label{unitarity}
 |0^{\prime}\rangle &\leftarrow& \cos\alpha |0\rangle + e^{i\beta'}\sin\alpha|1\rangle \cr
 |1^{\prime}\rangle &\leftarrow& e^{-i\beta'}\sin\alpha|0\rangle - \cos\alpha |1\rangle,
\end{eqnarray}
\noindent which is obviously a unitary transformation --rotation
in the Bloch sphere defined by angles $(\alpha,\beta')$-- for the B
basis $\{|0\rangle,|1\rangle\}$ in the range $\alpha \in [0,\pi]$
and $\beta' \in [0,2\pi)$. After some cumbersome calculations, it
turns out that the expression  for a minimum discord $\Delta$ of
the Introduction exhibits a positive and nonsingular Hessian, 
convex for the relevant range of values of $(\alpha,\beta')$. Our
expression possesses thus a unique global minimum, that occurs when
the concomitant partial derivatives vanish. This happens whenever
we have $(\sin\alpha=\frac{\sqrt{2}}{2},\sin\beta'=0)$.

\nd The present results correspond to pairwise
entanglement and quantum discord for the infinite $XY$ model at any
temperature, including zero-one. This implies that one does not
really need  to ``solve" the model in the sense of sufficiently
augmenting the number of spins in the chain for the results to be
thermally relevant.  $T$ here is an  {\it actual},
thermometer-measurable temperature, since we are tackling a
``real" thermodynamic system. This is to be confronted to the vast
XY-literature associated to finite spin-numbers, where $T$ is not, strictly speaking, 
well-defined in the thermodynamics sense.

\nd A comparison between the discord Qd and the entanglement of
formation $E$ at $T=0$ is displayed in Fig. 1  (from now on we
shall take the Boltzmann constant $k=1$). Qd and $E$ are
depicted versus the external magnetic field $h$ (anisotropy
$\gamma=\frac{1}{2}$) for the nearest neighbor configuration
$R=1$. Remarkably enough, the Qd measure exhibits a maximum in the vicinity of 
the factorizing field $h_f=\sqrt{1-\gamma^2}$. Both Qd and $E$
seem to decay in the same fashion. The classical correlations (CC)
for the same configuration are depicted in the inset of Fig. 1.
Notice that all quantities here considered, i.e., Qd, $E$, or CC,
are ultimately described in terms of several $G_R$s for all
configurations, so that they all diverge at the QPT (for $h=1$) in
the same way.

\nd As an  illustration consider   the magnetization given by
$M_z(h)=\frac{1}{2}G_0=\frac{\partial}{\partial h}
\frac{1}{2\pi}\int_0^{\pi} d\phi [\gamma^2\sin^2\phi + (h-\cos
\phi)^2]^{1/2}$. For $\gamma=1$ we have
$M_z(h)=\frac{\partial}{\partial h} \big( \frac{2(h+1)}{2\pi} E
\big[\frac{2\sqrt{h}}{h+1} \big] \big)= \frac{1}{2\pi} \big[
\frac{h-1}{h} K \big(\frac{2\sqrt{h}}{h+1}\big) + \frac{h+1}{h} E
\big(\frac{2\sqrt{h}}{h+1}\big) \big]$, where $K(E)$ is the
complete elliptic integral of the first(second) kind. Since
$\frac{d}{dh}M_z$ diverges in logarithmic way at $h=1$,
 as also do   the divergence of $K$ and 
 the first derivatives
of $E$, Qd, and CC. In other words, they all signal the presence
of a $h=1$-QPT at zero temperature (except for the isotropic
case $\gamma=0$). In fact, the possibility of detecting a QPT at finite $T$ by using Qd 
has been recently considered by Werlang {\it et al.} \cite{Werlang}. They perfom an interesting
analysis of the role of the temperature and Qd in several quantum systems. We remember 
that a different concept such as nonlocality --as measured by the maximum violation of the well known
Clauser-Horne-Shimony-Holt Bell inequality-- was also considered as a QPT in \cite{Batle2010} (also in the
context of the XY-model). In the present work we do not focus attention on this particular issue of QPTs, but study
instead the comparison between entanglement and quantum discord for finite and infinite systems at non-zero
temperature.

Fig. 2 depicts the the same quantities as  Fig. 1 for several
configurations. As we increase the relative distance from $R=1$ to
2, 3, and $\infty$, the corresponding Qd's diminishes and also decays
in faster and faster fashion with $h$. Notice that while  entanglement (not
shown here) globally diminishes,  Qd only tends to vanish
for $h>1$ and $R=\infty$. The inset here depicts the CC for the same
configurations. They decreasing in the same fashion. While $E$
tends to zero, both Qd and CC remain nonzero, regardless of the
distance between spins along the infinite chain.

 As soon as we introduce a non-zero temperature things
drastically change. In Fig. 3 we display several quantities at
different temperatures ($T=0.01,0.1,0.3,0.5,1$): $R=1$ and
$\gamma=\frac{1}{2}$. Fig. 3(a) depicts the entanglement of
formation $E$ for  states $\rho_{ij}^{(R)}$ (\ref{rho2}) versus
the magnetic field $h$ as we increase the temperature. $T$ lowers
and broadens the region of null entanglement from a point at the
factorizing field $h_f$ ($T=0$) to finite intervals centered at
$h_f$. Eventually, $E$ becomes finite at higher values of $h$.
This temperature-generated entanglement is depicted quantitatively
in Fig. 3(b), where the region of zero entanglement extends from a
point at zero $T$ to a finite-sized region as $T$ grows. The
aforementioned region ceases to be finite beyond a critical
temperature that depends on the particular $R$'s and $\gamma$'s
involved therein. We discern some resemblance with a  phase
diagram: within  the area encompassed by the two curves
of Fig. 3(b) no entanglement is detected.  It is
surprising  that, for the whole region, Qd globally
diminishes and tends to be concentrated in the null-$E$ region, as
 can be seen in Fig. 3(c). These facts allow one to  readily appreciate how different
 is the behavior of entanglement vis-a-vis that of Qd.
 The role of classical correlations can be observed in  Fig. 3(d).
 For the same set of temperatures employed above
  CC decreases  i) as
we augment $T$ and ii) for increasing values of $h$, a  behavior
 different from that  of  entanglement: while CC never vanishes,
it is larger  wherever $E=0$. Both $E$ and CC coexist for high values of $h$.
We are dealing with a system for which, as we
increase the temperature, entanglement survives --although barely-- for high values of $h$. This fact clearly affects the
existence of finite discord- or CC-values.  Recall that this was
the case already at $T=0$. The role of the factoring field $h_f$
in defining higher or lower values of Qd becomes crucial. To further
analyze the nontrivial relation between entanglement $E$ and
quantum discord Qd at finite $T$ it would enlightening to consider
a physical system for which $E$ would increase with the
temperature. Such is the Heisenberg model's scenario,
also a statistical mechanical model used in the study of critical
points and phase transitions of magnetic systems
\cite{baxter,Arnesen}.

\section{Two qubits in the (finite) Heisenberg model}

\nd First of all note that because of its finitude the  system  is not immersed in an
infinite thermal bath. Thus, we cannot stricto-sensu speak of a
``temperature''. However, the results to be presented are illustrative of the
intricacies of entanglement and quantum discord. Following the
interesting work of Arnesen {\it et al.} \cite{Arnesen}, we
concern ourselves with the issue of {\it thermal entanglement} but
extend the discussion so as to encompass {\it thermal discord}.
The Hamiltonian for the 1D Heisenberg spin chain with a magnetic
field of intensity $B$ along the $z$-axis reads

\begin{equation} \label{hamiltonian}
H\,=\,\sum_{i=1}^{N} (B\sigma^{i}_{z} \, + \, J_H \vec \sigma^{i} \,
\vec \sigma^{i+1}),
\end{equation}
\noindent where $\sigma^{i}_{x,y,z}$ stand for the Pauli matrices
associated to the spin $i$. Periodic boundary conditions are
imposed ($\sigma^{N+1}_{\mu}=\sigma^{1}_{\mu}$). $J_H$ is the
strength of the spin-spin repulsive interaction (only the
anti-ferromagnetic ($J_H>0$) instance is discussed). If we limit
ourselves to the case $N=2$, we  deal with two spinors, i.e., with
a two-qubits system. So as to speak of ``thermal equilibrium" we
consider  the thermal state \cite{Arnesen}

\begin{equation}\label{rhot}
\rho(T)=\frac{\exp(-\frac{H}{k_{B}T})}{Z(T)},
\end{equation}
with $Z(T)$  the
partition function. Expressing both $H$ and $\rho(T)$ in the
computational basis $|00\rangle,|01\rangle,|10\rangle,|11\rangle$
we obtain

\begin{equation}
H = \left( \begin{array}{cccc}
2J_H+2B & 0 & 0 & 0\\
0 & -2J_H & 4J_H & 0\\
0 & 4J_H & -2J_H & 0\\
0 & 0 & 0 & 2J_H-2B \end{array} \right).
\end{equation}
After defining, for convenience's sake, \newline \noindent
$e_{wmy}=\exp{\,(-2w-2y)};$
\newline \noindent $e_{wp}=\exp{\,(-2w)}+\exp{\,(6w)};$\newline \noindent
$e_{wm}=\exp{\,(-2w)}-\exp{\,(6w)}$; \newline \noindent $e_{wpy}=
\exp{\,(-2w+2y)},$ \newline \noindent with $w=J_H/k_{B}T$ and
$y=B/k_{B}T$, we also get

\begin{equation}
\rho(T) = \frac{1}{Z(T)}\left( \begin{array}{cccc}
e_{wmy} & 0 & 0 & 0\\
0 & e_{wp}/2 & e_{wm}/2 & 0\\
0 & e_{wm}/2  & e_{wp}/2 & 0\\
0 & 0 & 0 & e_{wpy} \end{array} \right),
\end{equation}
\noindent  In this case the concurrence of $\rho(T)$ reads
\cite{Arnesen}
\begin{eqnarray}
C&=&0;\,\,\, \,\,\,\,\,\,\,\,\,\,\,\,\,\,\,\,\,\,\,\,\,
\,\,\,\,\,\,\,\,\,\, \,\,\,\,\,\rm{for}\,\,\, T \ge T_c, \cr C&=&
\frac{e^{8w}-3}{1+e^{-2y}+e^{2y}+e^{8w}}; \,\,\rm{for}\, T <
T_c,\end{eqnarray} \noindent Recall that there is no entanglement
beyond a certain critical temperature $T_c=8J_H/(k_B\ln 3)$
\cite{Arnesen}, as can be seen from the previous $C-$computation.
 Also, there is a change in the structure of the ground state of
hamiltonian (\ref{hamiltonian}) when the magnetic field reaches
the critical value $B_c=4J_H$. In the limit of zero temperature,
the ground state of the system may be represented by three
different pure states:  i) for $B<B_c$ (non-degenerate), the
thermal state reduces to the singlet state
$|\Psi^{-}\rangle\langle \Psi^{-}|$,  ii) at $B=B_c$ (two-fold
degenerate) $\frac{1}{2}|\Psi^{-}\rangle\langle
\Psi^{-}|+\frac{1}{2}|11\rangle\langle 11|$, iii) whereas for
$B<B_c$ (non-degenerate) we have $|11\rangle\langle 11|$. The
previous $B-$distinctions are crucial in order to understand how
the concomitant thermal state will respond  to $T-$changes. We
expect accompanying behaviors from  $E$ and Qd whenever the
initial state is pure  (both quantities coincide in such case).
Differences should emerge  for $B>B_c$ whenever we study the
unexpected behavior of increasing $E$ versus $T$ as far as Qd is
concerned. The computation of Qd is in the present case  analytic
and corresponds to $\sin\alpha=\frac{\sqrt{2}}{2}$ for any
$\beta$. The pertinent  scenario is the subject of Fig. 4 (let us assume $J_H=1$,
so that $B_c=4$). For the $B$-range of values that are smaller than the critical value $B_c$,
entanglement, discord and CC all diminish as the $T$ increases, as shown in Fig. 4(a).
This behaviour also occurs at $B=B_c$ and is  depicted in Fig. 4(b). Notice in both cases the sudden death
of  entanglement, whereas the other quantities ``survive'' in indefinite fashion. 
Fig. 4(c) plots the same quantities for a magnetic field $B>B_c$. Remarkably enough, in this case entanglement as well as the quantum
discord {\it augment} as $T$ increases. In this case, again,  $E$ suddenly vanishes while the persistence of the quantum discord and
CC. [see Fig. 4(d)] stresses the fact that for $B$ significantly differing from $B_c$, $E$ is minimalfor all $T$ while the quantum discord survives. 

Overall, entanglement and quantum discord display  similar behaviours --although with clear differences-- for a finite
quantum system [two spins in the Heisenberg model] but become radically different from each other when we consider a system
in  the thermodynamic limit (such as the XY-model).

\section{CONCLUSIONS}

We have compared entanglement $E$ and quantum discord Qd for magnetic systems at finite temperatures, comparing their behavior with that of classical correlations as well. 
 It is clear that, some similarities notwithstanding, $E$ and Qd behave in quite different fashion in the thermodynamic limit. The distinction we are trying to establish here is 
blurred in the case of finite systems. We conclude that for realistic systems $E$ and Qd should both be studied in independent fashion, as they reflect on differen aspects of the
 quantum world.

\vskip 3mm {\bf ACKNOWLEDGEMENTS} \vskip 3mm
J. Batle acknowledges fruitful discussions with J. Rossell\'{o} and M. del M. Batle.
M. Casas acknowledges partial support under project FIS2008-00781/FIS (MICINN) and FEDER (EU).

\begin{figure}[htbp]
\begin{center}
\includegraphics[width=11cm]{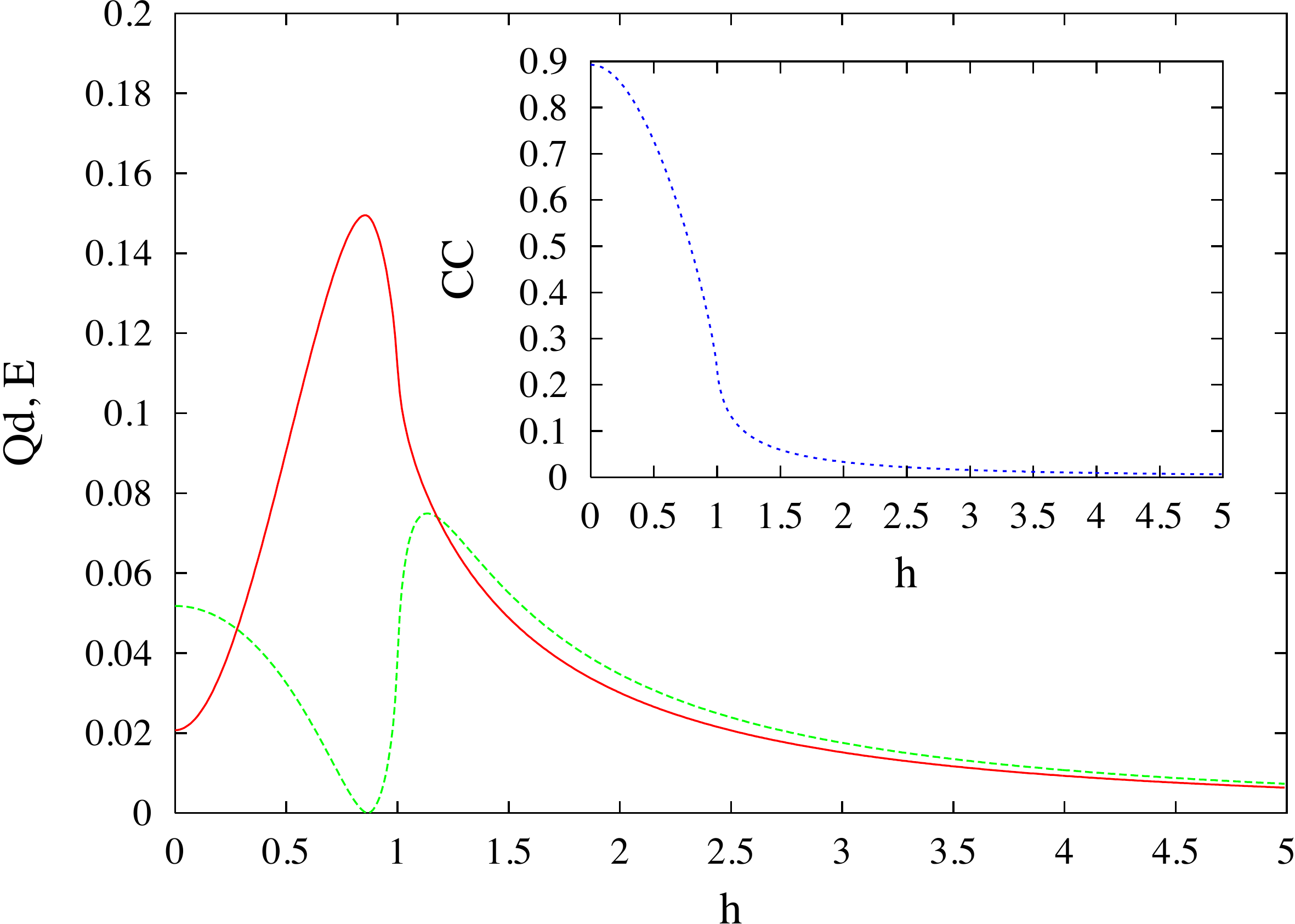}
\caption{Color online) Plot of quantum discord Qd (upper solid curve) and entanglement of formation E (lower dashed curve)
vs the external magnetic field $h$ for two qubits in infinite the $XY$ model (nearest neighbors),
with anisotropy $\gamma=\frac{1}{2}$ at T=0. The region around the factorizing field $h_f$ concentrates maximum Qd. Inset
depicts the corresponding classical correlations CC vs $h$. See text for details.}
\label{fig1}
\end{center}
\end{figure}

\begin{figure}[htbp]
\begin{center}
\includegraphics[width=11cm]{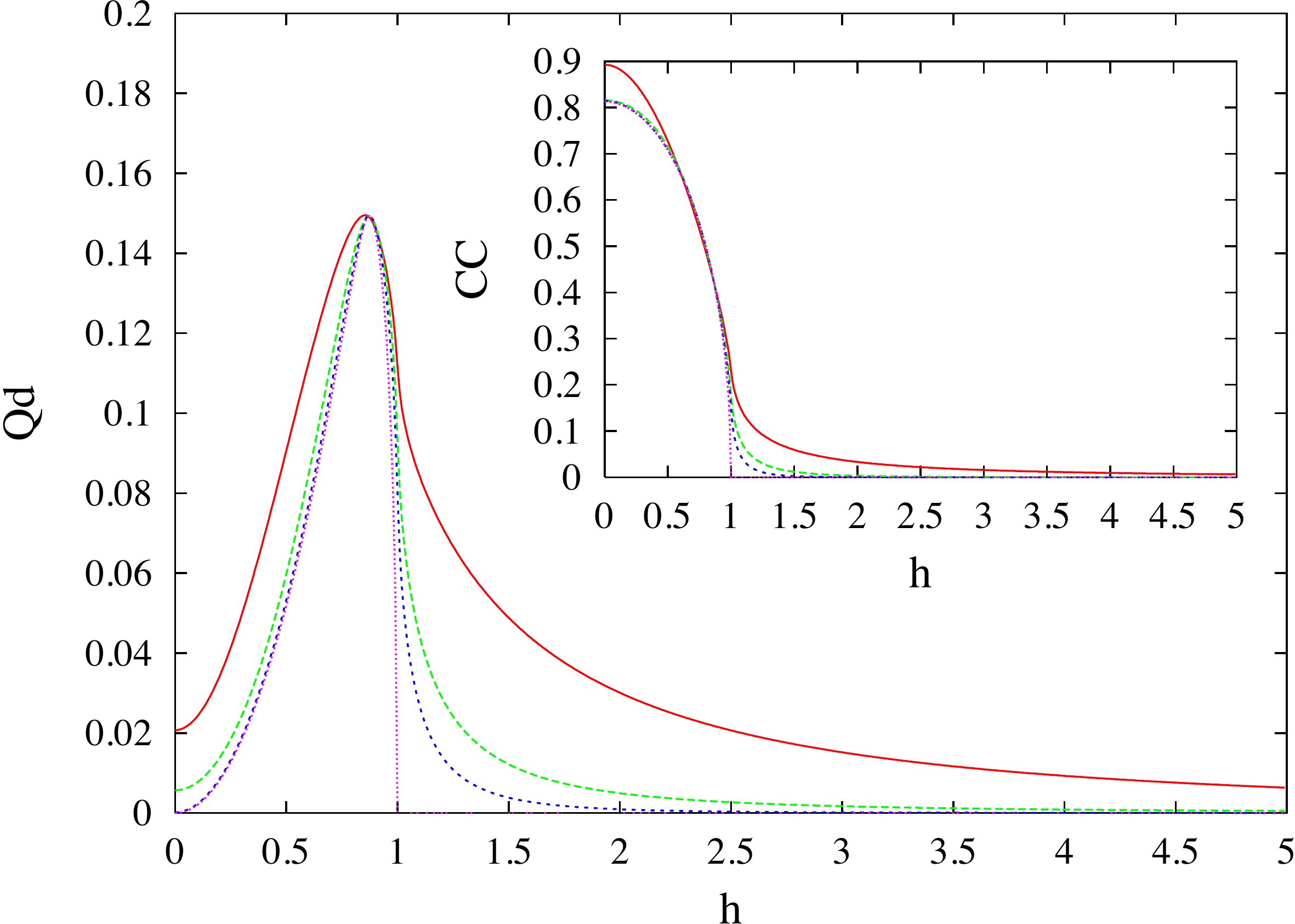}
\caption{(Color online) Plot of Qd for the same settings as in Fig. 1 for different relative distances
$R=1$ to 2, 3 and $\infty$ between spins. The further they are separated, the more they collapse into
a single curve, which is zero for $h>1$. Notice that E rapidly tends to zero for all $h$ in the limit
$R \rightarrow \infty$, while the corresponding Qd remain finite. A similar behavior occurs for CC as depicted in
the inset. See text for details.}
\label{fig2}
\end{center}
\end{figure}

\begin{figure}[htbp]
\begin{center}
\includegraphics[width=11cm]{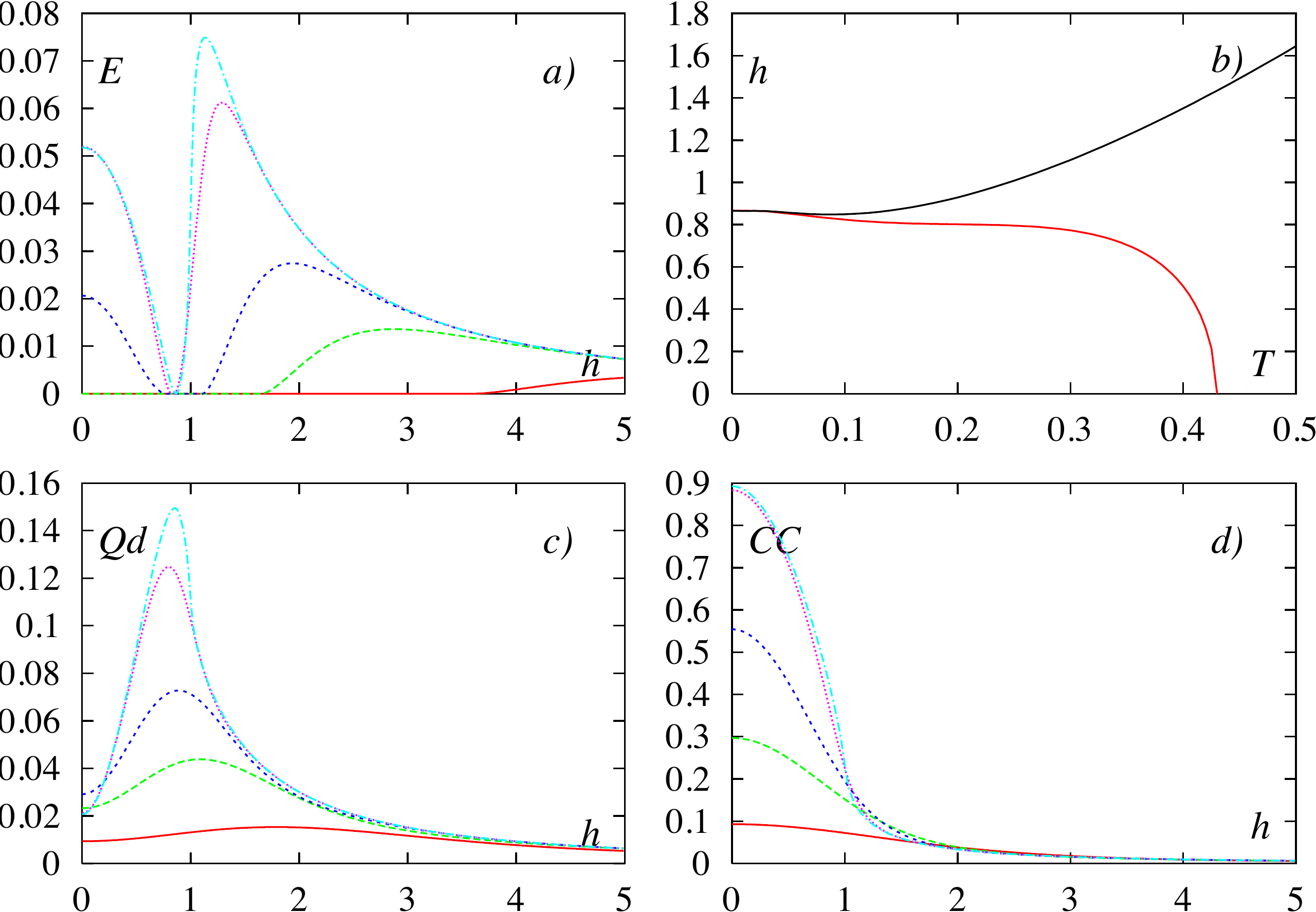}
\caption{(Color online) (a) Value for E vs $h$ for finite temperatures T=0.01,0.1,0.3,0.5,1 (from top to bottom) for $R=1$ and
$\gamma=\frac{1}{2}$. Notice how
the region of null entanglement spreads form a point (at the factorising field $h_f=\sqrt{1-\gamma^2}$) to a region. (b) Plot
of the aforementioned region of zero entanglement. The upper and lower curves define de limits of $h$ for a given $T$ where null E is found.
This figure resembles a phase diagram-like plot where the regions of zero and nonzero entanglement are defined.
(c) Qd exhibits a particular behavior as T increases which tend to be maximum within the limits of zero E. (d) CC vs $h$ plot
for the same temperatures. An overall decreasing tendency is apparent. See text for details.}
\label{fig3}
\end{center}
\end{figure}

\begin{figure}[htbp]
\begin{center}
\includegraphics[width=11cm]{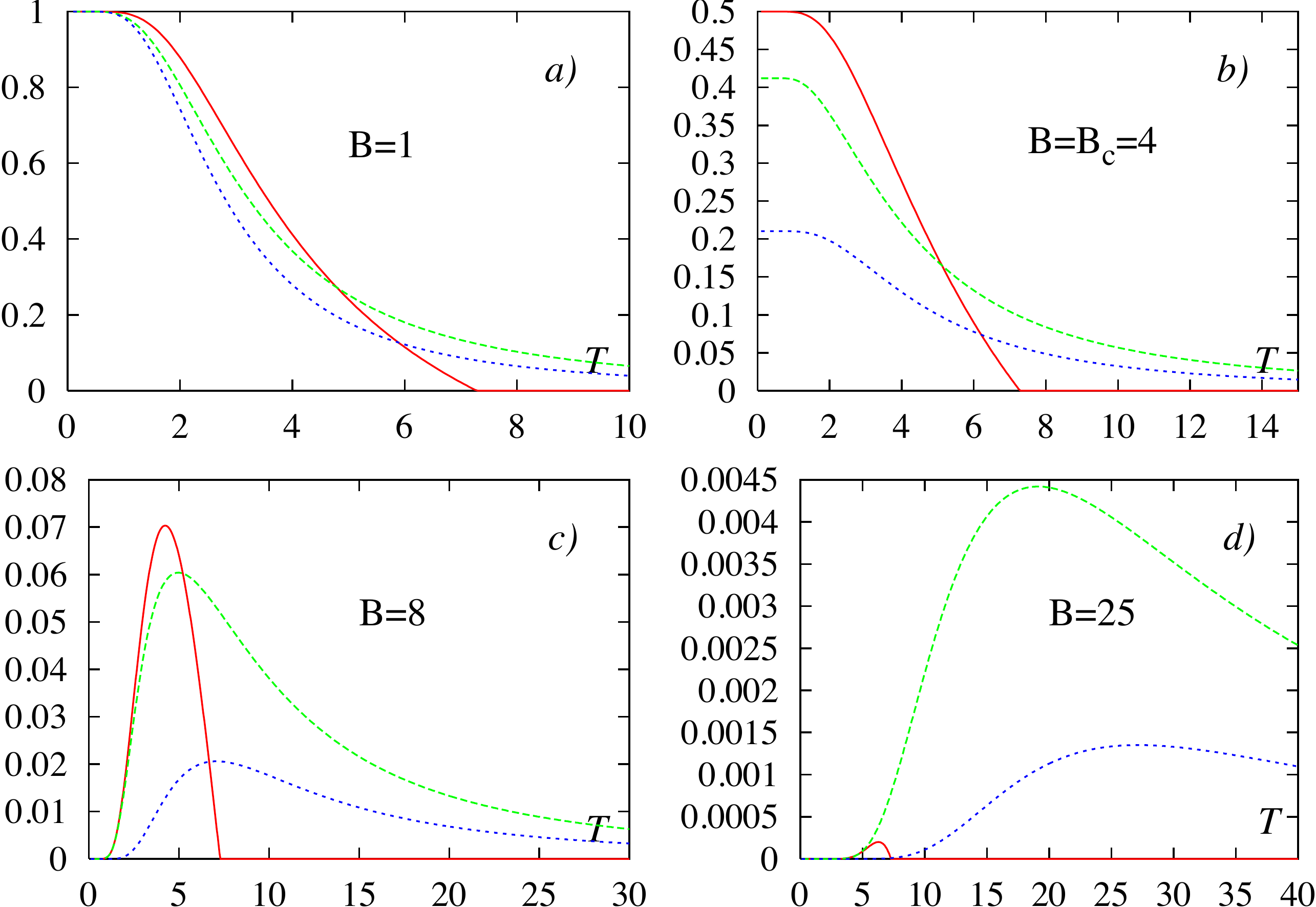}
\caption{(Color online) (a) E (solid line), Qd (long-dashed line) and CC (short-dashed line) vs T plots for the thermal state of
two qubits in the Heisenberg model at the magnetic field $B=1$ ($B<B_c$). (b) Identical plot for $B=B_c=4$. (c) Identical plot for
$B=8$ ($B>B_c$). (d) Plot of the previous quantities for a high value of $B$ ($B=25$). See text for details.}
\label{fig4}
\end{center}
\end{figure}


\begin{thebibliography}{}
\bibitem{qpt} Recall that a QPT is a qualitative change that occurs
in the ground state of a many-body system due to modifications
either in the interactions among its constituents or due to the
effect of an external probe.
\end{thebibliography}
\end{document}